# How to Promote Informal Learning in the Workplace?
## *The Need for Incremental Design Methods*

Carine Touré[1,3], Christine Michel[1] and Jean-Charles Marty[2]

[1] *INSA de Lyon, Univ Lyon, CNRS, LIRIS, UMR 5205, F-69621 Villeurbanne, France,*
[2] *Université de Savoie Mont-Blanc, CNRS, LIRIS, UMR 5205, F-69621 Villeurbanne, France,*
[3] *Société du Canal de Provence, Le Tolonet, France*
{carine-edith.toure, christine.michel, jean-charles.marty}@liris.cnrs.fr

Abstract: Informal Learning in the Workplace (ILW) is ensured by the everyday work activities in which workers are engaged. It accounts for over 75 per cent of learning in the workplace. Enterprise Social Media (ESM) are increasingly used as informal learning environments. According to the results of an implementation we have conducted in real context, we show that ESM are appropriate to promote ILW. Indeed, social features are adapted to stimulate use behaviors and support learning, particularly meta-cognitive aspects. Three adaptations must nevertheless be carried out: (1) Base the design on a precise and relatively exhaustive informational corpus and contextualize the access in the form of community of practice structured according to collaborative spaces; (2) Add indicators of judgment on the operational quality of information and the informational capital built, and (3) Define forms of moderation and control consistent with the hierarchical structures of the company. Our analysis also showed that an incremental and iterative approach of user-centered design had to be implemented to define how to adapt the design and to accompany change.

## 1 INTRODUCTION

Lifelong learning is an approach to education that has been addressed since the 1970s to provide the skills and knowledge needed to succeed in a rapidly changing world (Sharples, 2000). It includes formal, non-formal and informal learning (Commission of the European Communities, 2000). Unlike informal learning, formal and non-formal learning are structured with tools or training sequence. The latter occurs during daily experiences, while working or interacting with other people. It is characterized by the merger of learning with the everyday work activities in which workers are engaged (Longmore, 2011) and is motivated by personal needs. Informal learning is of central importance for enterprise since it accounts for over 75 per cent of learning in the workplace (Bancheva and Ivanova, 2015). It is the most important way to acquire and develop skills required in professional contexts.

The Knowledge Management (KM) research field promotes the management and maintenance of knowledge sharing in the workplace. Three generations of technologies were privileged for informal learning (Ackerman *et al*, 2013; Hahn and Subramani, 1999). Two main strategies can be identified to manage knowledge: valuation of informational capital and valuation of human capital with collaboration (Ackerman *et al*, 2013; Wenger, 2000).

The first generation considers that workers can continuously learn and be able identify solutions to problems they can meet during working activities. They have to look for information on processes and know-how related to their activity. To support them, enterprises produce relatively exhaustive information corpuses on working activity and make them accessible. Despite their exhaustiveness, these knowledge databases remained most of the time unused because they were maladjusted to collaborators needs and characteristics; particularly regarding information access and training (Hager, 2004; Graesser, 2009). Moreover, access tools to this information are not dedicated to learning process. Indeed, Graesser (2009) recommended to privilege training objectives based on auto-regulation and meta-cognition ; and by this way help learners to "learn how to learn". He describes (Graesser, 2011) various principles based on fun, feedback or control to support learning.

The second generation focus was on expertise sharing and identification of experts able to provide useful information to collaborators. Communities of practice (CoP) were commonly adopted by enterprises to help practitioners express, share and exploit their knowledge (Pettenati and Ranieri, 2006; Wenger, 2000). Direct interaction between peers was recognized to facilitate knowledge transfer and improve information quality (Wang, 2010). However, the lack of information completeness, accuracy in identification and recommendation of expert, privacy protection and control revealed some limits (Ackerman *et al*, 2013). CoPs have remained hardly ever used.

The third generation combines principles of both first and second generations. It is characterized by collaborative information spaces merging information repositories, communication and collaboration processes. Many enterprises chose to implement enterprise social media (ESM) to improve organizational performance, especially in the knowledge sharing context (Ellison, Gibbs and Weber, 2015). They integrate management of working activity, knowledge management strategies and social aspects promoting interactivity between peers (Dennerlein *et al*, 2015; Leonardi, Huysman and Steinfield, 2013; Riemer and Scifleet, 2012). ESM foster informational and social capital valuation; they are particularly well adapted to find and interact with collaborators, receive and seek for help (Ackerman *et al*, 2013). They are also easier to manipulate, more attractive and interactive than traditional collaborative environments. They fulfill users' needs for usefulness and gratification (Ersoy and Güneyli, 2016). Indeed, they allow the recognition of each one in the contributions made and permit social connections materialized by simple actions as following a post or as commenting. Nevertheless, the free access to information, contribution and cooperation features has opened the door to misuse leading to a lack of efficiency in the exploitation of information resources or a feeling of harassment (Turban, Bolloju and Liang, 2011).

Our objective is to study to what extent ESM are actually adequate tools to implement informal learning strategies. More specifically, we will study what social features are the most effective to match learning objectives stated by Graesser and how to make them coherent with the objectives and practices of the organization and collaborators. The long-term objective is to favor a sustainable use. To answer these questions, we present in the next section ESM characteristics and how they can significantly support informal learning in the workplace. This helped us identify various design propositions. We implemented these propositions in a real context to evaluate their accuracy and refine them. This study is presented in the third section of the paper.

## 2 USING ESM FOR INFORMAL LEARNING

### 2.1 Pros

ESM features promote construction and identification of relevant information. Comments within social media are an emblematic form of expression and a communication tool for users to effectively judge the quality of information and easily participate to content construction. Indeed, information captured within informal learning tools evolves and may become rapidly outdated. Comments have the advantage that workers can communicate and participate online to the construction of the knowledge corpus (Kaplan and Haenlein, 2010), they reduce the risk of forgetting or losing practice. Appreciations left by users provide us with an additional way to evaluate information quality and to promote information submission. They can be formalized as in some wikis where content posted can be qualified with completeness and readability indicators. These indicators allow collaborators to form their opinion on the content and better understand how they can participate to its refinement. This feedback helps authors to be aware of the usefulness of their publications (Kietzmann, Hermkens andMcCarthy, 2011) and helps to build their reputation. Moreover, wikis frequently use these features to support collaborative innovation, problem resolution and more generally help organizations improve their business processes (Turban, Bolloju and Liang, 2011).

ESM provide visibility and persistence of several communicative actions like download, content publication, identification of what others do, status update, profile creation (possibilities to highlight particular aspects of themselves), connecting with or following people (Leonardi, Huysman and Steinfield, 2013; Stocker and Müller, 2013). They expand (and precise) the range of people, networks and contexts from which people can learn across the organization.

Making communicative activities visible also allows self-regulation. Notifications, number of appreciations, new submissions, etc. help identify what and how others do, evaluate what one's do and adjust one's own behavior. It promotes meta-cognition and meta-knowledge (learn how to learn) (Schön, 2000). This awareness thus becomes an intrinsic motivator to construct one's own numerical identity through indicators (Zhao, Salehi and Naranjit, 2013). Being involved in a group helps collaborators develop meta-social knowledge and facilitates their ability to collaborate and coordinate (Janssen, Erkens and Kirschner, 2011), particularly within CoPs.

## 2.2 Cons

Janssen *et al* (2011) identifies two groups of risks linked to the problems of acceptance and to the use of ESM and quality of content published by collaborators.

The acceptance and the ability to use play a basic role on the initial and continuous use of technologies. The process of acceptance begins with the construction of initial beliefs towards the information system. They are generated by external stimuli such as system quality, service quality, knowledge quality or information quality (DeLone and McLean, 2003; Jennex and Olfman, 2006; Kulkarni, Ravindran and Freeze, 2007; Venkatesh *et al*, 2003). These beliefs are moderated by personal factors like age, previous experience or service quality (Venkatesh, Thong and Xu, 2012). They also influence the ease of use of the system. Indeed, an efficient use may require high levels of literacy and technical proficiency in seeking for information, evaluating its usefulness and truthfulness or connecting with remote people or computers (Benson, Johnson and Kuchinke, 2002; Turban, Bolloju and Liang, 2011). Contextual characteristics of collaborators are most of the time not considered during the design process (Longmore, 2011). To develop meta-social skills and improve communication as proposed by Graesser (2011), users need clear learning objectives and awareness on peers feedback and information quality. They also need recognition of what they do (improvement of professional reputation, acknowledgement from community, being informed that their actions are appreciated by others) (Wang and Noe, 2010). Moreover, policies and structure of governance (i.e. monitoring, control or filtering of system accesses) have to be established as well as management campaign of training) (Turban, Bolloju and Liang, 2011). These solutions are money and time consuming, especially for limited IT budgets and companies that seek rapid and simple collaborative solutions. After this initial cycle of use, the user acquires an experience that helps him to construct new beliefs and experience confirming or refuting the previous ones ; this impacts his attitude towards the system (satisfaction or dissatisfaction) and his intention to use the system (Bhattacherjee, 2011; Bhattacherjee, Limayem and Cheung, 2012).

The second group of risks concerns the validity and quality of information created and published. Despite the fact that published content is most the time not anonymous within ESM, it can be useless for informal learning since information is often poorly detailed and proofread, particularly if knowledge objects manipulated are of technical nature. Within social media, posts are very often brief and people give generic information without giving details. This may be suitable for updates, but not for the construction of the core information corpus. Moreover, people may engage in informal behavior when using social media. Activities like using improper language, publishing information that is confidential, using incomplete information or using ratings or comments to harass colleagues may be common. The ability to discern the quality of the accessible information is mostly incumbent upon users and they have little control in these environments, which is one principle of social media (Bhattacherjee, 2011; Turban, Bolloju and Liang, 2011). These risks may negatively affect the social and learning environment and call into question the expected learning processes.

## 2.3 Summary and Proposition

ESM appear to be suitable to support informal learning in the workplace. They supply functionalities that promote and facilitate collaboration, knowledge sharing, user motivation and visibility, and information persistence. They also propose reflexive indicators that facilitate the analysis and coordination of collective activities, social connection and learning. These characteristics position workers and their needs at the heart of the learning environment, making ESM appropriate tools to support informal learning in the workplace. However, their use may be inefficient due to the profile of workers, who are adult learners and need to be aware of the value of their participation in the learning group: they seek concrete personal and professional feedback, usefulness and gratification. Moreover, the quality of information published may be problematic regarding learning strategies.

To reduce the risks related to information quality, we believe that it is important to base the learning environment on a precise and exhaustive information corpus. This informational architecture which is most of the time already formalized into information systems can then be enriched with collaborative characteristics. Literature review showed that indexation and structuration of information have to be reviewed to facilitate contextual access. A search engine and indexation tags are fundamental elements to guaranty a transversal access to information.

Activity's contextual aspect such as the one proposed by CoPs can be reproduced with structured wikis according to enterprise's working communities. Various elements have to be considered to guaranty quality and trustworthiness of published content; and also facilitate contribution: select useful information, organize it according to specific template files, and organize validation according to hierarchical decision-making structures of the community.

As regards to learning support, literature review showed three additional characteristics of ESM to promote users' engagement: visibility and reflexivity. Comments and appreciations (*e.g.* "Likes") can be considered as tools for expression and communication, allowing collaborators to provide feedback and participate to construction of contents. These features allow them to be involved into the co-construction of knowledge and maintain an updated available information which is important for the quality of learning processes. Awareness indicators like notifications (of new submissions, who and when, number of comments) promote the construction of meta-cognitive skills for self-regulation and stimulate participation. Indicators of information quality facilitate identification of useful content and collaboration by a critical analysis of items to be added to update and improve contents.

Finally, to minimize risks of misuse of the environment, we propose to use a user-centered, incremental and iterative design methodology. This methodology allows to identify characteristics and preferences of users and to design a contextually adapted environment. The incremental and iterative nature of the approach also makes it possible to accompany the change associated with the introduction of a new information system and thus to positively influence its acceptance and its initial and continuous use. Indeed, since informal learning is inherent in the employee's will and not stimulated by accompanying strategies, this characteristic appears fundamental. Analysis of core acceptance of technology models in the workplace showed that the acceptance model can be represented with a spiral (see Figure 1) structured with conditions of use. Every loop builds an artifact increasingly adapted to users' needs and behavior. We posit that the sustainability of our process can be effectively ensured by providing users with an artifact matching their profile and needs at each stage of this cycle.

We implemented our methodology in a real context. The objective was to identify the most adapted ESM features that promote informal learning, to assess the feasibility of the methodology and to identify a structuring order of the various items which have to be considered at each stage. We present the results of this experimentation in the next section.

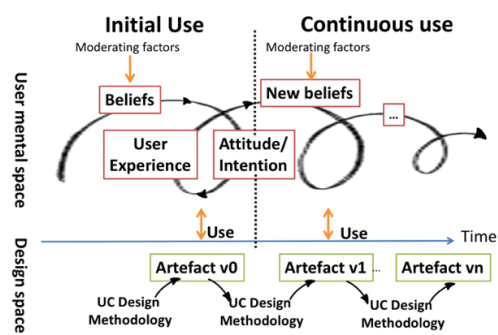

Figure 1: Incremental and iterative design of information systems for informal learning.

## 3 IMPLEMENTATION

### 3.1 Context and Constitution of the Working Group

The *Société du Canal de Provence* (SCP) is located in the south of France and specializes in services related to the treatment and distribution of water for companies, farmers and communities. The intervention territory is divided into ten geographic areas called Operating Centres (OC). Each OC corresponds to a community of practice in which we find three positions: the Operator (O), the Coordinator Technician (CT) (an operator who also has the role of manager of the community), and the Support and Customer Relationship Technician (SCRT). They are the responsible people for the maintenance of hydraulic infrastructures (canals, pumping stations, water purification stations, etc.). The operators need a wealth of knowledge about their work: there is a lot of (sometimes dynamic) information to learn and knowledge sharing is especially important.

To assist them, SCP produced in 1996 a knowledge book about the processes and hydraulics infrastructures. This information was accessible through a tool named ALEX (*Aide à L'EXploitation*).

It gathered information from returns on experience sheets developed in HTML format, and stored it within a directory on a dedicated server in each OC. Throughout its twenty years of existence, it was hardly ever used despite the fact that collaborators agreed with the learning environment principle. One main reason was because accessibility to information was not adapted. ALEX was a typical sample of traditional KM strategies based on knowledge books and produced in the 1990's. It is an appropriate context to work on means capable of supporting lifelong learning.

Four OCs were selected by the responsible person for the project to act as pilot OCs. Eleven employees coming from those OCs were invited to freely participate in the working group. They were chosen according to their experience and different positions, thus being representative of various trades within the company, and according to their use of the previous version of the knowledge book. The focus groups were moderated by ourselves and by a member of the working group (a board member, responsible for the ALEX project). We count in total twelve sessions conducted on a two years period. The first year consisted in formalizing the basic users' needs. Six meetings, separated by about two to three weeks, allowed us to propose a solution increasingly refined until a last version fully usable in the work context. The platform was made available to users for three months. At the end of this first year, a debriefing meeting was held on the eligibility of the proposed solution and a new analysis and design cycle was initiated. It took seven months and six working sessions.

## 3.2 The First Design Cycle

Results of the first cycle are presented more in detail in (Touré *et al*, 2015). In summary, this stage showed that the main requirement that emerged from the meetings was to propose easier ways to search for, submit and access knowledge (see Figure 2 zone 1,5,7) organized in collaborative CoPs according to the different OCs. The discussions allowed us to identify the general structure of navigation and organization of the information of the website and the methods of structuring knowledge, in particular eleven different structures of data sheets. A work of harmonization of the architecture of the various IS was carried out to integrate Alex with the other IS and with the intranet of the company. The objective was to facilitate the navigation between the different tools and thus their accessibility from every workstation and in mobility. A simplified numerical space reproducing a word processor office suite and various document templates were designed. Four user roles were proposed to control submissions and guaranty information quality – the reader, the contributor, the validator and the manager. The working group was in charge of attributing the different roles. For example, the validator roles were attributed to CTs who are responsible for each OC while the manager roles was attributed to ALEX project responsible person.

After a three months use, an evaluation was conducted and showed that this new version of ALEX match the basic users' expectations but lacked attractive items to guaranty a long-term usage (Touré *et al*, 2015). The second design cycle allowed us to work on these elements.

## 3.3 The Second Design Cycle

Discussions were about design of items for stimulation, control and monitoring of activity. They revealed two emerging groups of needs for readers/contributors and for validators/managers. The first ones were sensitive to the addition of social features and activity indicators (comments, ratings, notifications…) while the latter expressed expectations about monitoring activity via an activity dashboard. We present in the following subsection results related to social features, since the dashboard is still being developed currently.

### 3.3.1 Comments and Appreciations

Discussions on comments and appreciation were based on mockups presenting interactions that mimic what is commonly done in Web 2.0 knowledge construction tools like blogs or wikis: comments and "Likes" counting number of positive appreciations.

All the participants agreed with the idea of using comments as they are simpler means of communication than emails. They also make the sheets interactive, as they can be seen as an 'annotation tool'. However, they noted that contributors must be informed when a new comment is added on their experience sheet. Moreover, unlike comments left within classic social networks, SCP collaborators asked for moderation and archiving of comments to improve their readability and to control potential excess or harassment in relation to co-workers. Validators (collaborators with enough expertise who are in charge of electronic validation of experience sheets) will manage and ensure that propositions made within comments are effectively taken into account for the improvement of sheets. They are also in charge of archiving comments.

The 'Like' functionality gave rise to much discussion. Some participants had concerns about the real meaning of the term 'like', potential abuse (if a 'Like' is just given by affinity and does not reflect the quality of a contribution) and the negative impact it could have on contributors' motivation if they do not receive any. Some participants thus asked for clarification by relabelling the functionality to 'useful sheet'. Others were very enthusiastic about it as they are already familiar in other social networks and consider it as 'playful' in a professional context. During the next session, where the resulting feature was shown to users, they finally argued that the 'Like' functionality, in the context of SCP, is not a key motivator for contribution but rather signifies the reactivity of other collaborators and the awareness of their feedback, the feeling of being in a human community that works. Ultimately, they agreed to consider 'Likes' as assessments of the sheet's content usefulness expressed by readers and to leave the term 'like' as is. Some adaptations have however been requested: replace the raised thumb by a smiling emoticon, and initiate discussions, among collaborators, on this social functionality to prevent the risks of misunderstandings and abuse.

### 3.3.2 Activity Indicators

Several pieces of information were proposed as representative of reflexive indicators: notifications of new publications, authors and date of submission, last sheets read, view of contribution status, number of comments received on a sheet. The view of publications and number of comments did not trigger any discussion as they have been already discussed in previous sessions (see Figure 2 zone 3,4). Notifications of new contributions published or consulted were mentioned to facilitate the identification of recent information and the interests of other collaborators. The identification of the actors, such as the last contributor or the last reader, was deemed useful for initiating direct discussions between colleagues. However, the identification of the successive contributors was not considered necessary, a validated form being considered as a collective work. The status of publications (pending, rejected, and accepted) has emerged due to the expressed need to know if and when the validator has taken into account a contribution. Finally, by considering possible use cases, the discussions revealed two ways of presenting these indicators: in a personal page linked to profiles (see Figure 2 zone 6 for access) and on COs front pages. The first page was seen as a way for each collaborator to follow his / her own activity and see its scope within the organization. The second was seen as a means for identifying the dynamics of a community, updated or useful information and thus initiating discussions among colleagues.

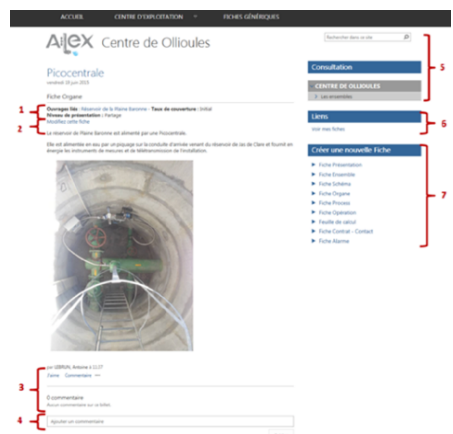

Figure 2: View page of a content form.

### 3.3.3 Information Quality Indicators

Three indicators were proposed to express information quality: readability, completeness with respect to the concept described and relevance (Lee, *et al.*, 2002). The objective is to inform the user of the reading effort necessary to realize the information presented in real situations of work or problem solving. There was general support for the use of such indicators. Discussions focused on evaluation scales, how values were allocated, and the names of indicators. To describe readability, participants proposed a 4 level scale: *operational* (the information on the sheet is immediately or quickly exploitable, such as alarms records specifically describing each step to perform a corrective maintenance operation); *support* (can be used in case of emergency but requires more analysis for information appropriation); *acquisition* (general information to train the reader); and *sharing* (information that needs further work). An agreement was reached on the term 'presentation level' to name the indicator. Completeness was found useful using the name 'Level of coverage'. The evaluation scale of this indicator is on three levels: weak, medium and good. This indicator was not deemed appropriate, as content is relevant if accepted for publication by the validator. As with the readability indicator, the completeness assessment of the sheet is made by the validators. The participants did not deem it useful to depict this indicator with an icon (stars, lights ...) and preferred the indications to be directly written in the header of each content form (see Figure 2 zone 1).

## 4 EVALUATION

### 4.1 Methodology

A qualitative evaluation has been done to measure the design quality and the potential learning effectiveness. Criteria for design quality are deriving from uses success factors identified in the TAM, UTAUT and ISSM models of technology acceptance (DeLone and McLean, 2003; Venkatesh *et al.*, 2012): Use (Use), Usefulness (Usef), Satisfaction (Sat), Percieved benefit (Ben), Usage intention (UI). Indeed, successful use of the tool is related to positive satisfaction, attitude and intention; this is why we focus on these criteria. We measured learning effectiveness according to users' statements on impact on use, work habits and performance (IOU&W).

ALEX with social functionalities was made available for four months. Ten collaborators have been interviewed about their uses and positions about new social functionalities. A first group (group 1) was composed by five of them (named P1 to P5) that had participated to the design working group, while a second group (group 2) was composed by five other people (named P6 to P10) who were not involved at all in ALEX design. Interviews were individual and lasted one hour per person. During the interview, an interface of ALEX was available to help participants to contextualise and refine their appreciations. The interviews were anonymously recorded and manually encoded to identify the parts of sentences, called utterances, corresponding to the different criteria. A positive (+), neutral (=) or negative (-) polarity was assigned to each selected utterance. An utterance was considered as neutral when participants said that they did not know how to answer a question or when it was not possible to detect a polarity in the given answer. We analysed participants' appreciations according to the number of statements and polarity on each criteria and compare the two groups to measure if the working group proposition are shared with the other collaborators.

### 4.2 Results

#### 4.2.1 General Results

111 utterances (n=111) were collected. Table 1 describes their distribution according to the six criteria and the three polarities in frequency and percentage. We note that appreciations are globally positive (60.2%). Only 11.3% are negative and 28.5% are neutral. Usefulness is the most expressed statement (40) and is globally positive (52.5%) even if one third of the participants don't have an accurate point of view about it (32.5%). Satisfaction and impact on work and performance are the most positive criteria (with respectively 90% and 72.7%). A third of the participants (35.7%) express positive statements about usage intention while nearly half of them (42.8%) have no real idea of the kind of usage they can introduce. Statements about real uses are diverse. Half of the participants express positive uses (50%) but a third of them (31.3%) didn't use Alex. Comments of participants related to each criteria presented in the section 4.2.3 are useful to refine and understand these results.

Table 1: Distribution of utterances according to criteria (frequency) and polarities (percentage).

|       | Use  | Usef | Sat | Ben | IOU&W | UI   | Means |
|-------|------|------|-----|-----|-------|------|-------|
| *n*   | *16* | *40* | *20*| *10*| *11*  | *14* |       |
| + (%) | 50   | 52.5 | 90  | 60  | 72.7  | 35.7 | 60.2  |
| - (%) | 31.3 | 15   | 0   | 0   | 0     | 21,4 | 11.3  |
| = (%) | 18.7 | 32.5 | 10  | 40  | 27.3  | 42.9 | 28.5  |

Table 2: Group 1 and 2 comparison.

|         |        | Use  | Usef | Sat | Ben | IOU&W | UI   |
|---------|--------|------|------|-----|-----|-------|------|
| Group 1 | + (%)  | 37.5 | 25   | 35  | 20  | 27.3  | 7.1  |
|         | - (%)  | 6.3  | 5    | 0   | 0   | 0     | 7.1  |
|         | = (%)  | 12.5 | 7.5  | 5   | 30  | 18.2  | 28.6 |
| Group 2 | + (%)  | 12.5 | 27.5 | 55  | 40  | 45.5  | 28.6 |
|         | - (%)  | 25   | 10   | 0   | 0   | 0     | 14.3 |
|         | = (%)  | 6.3  | 25   | 5   | 10  | 9.1   | 14.3 |

#### 4.2.2 Group Answers Comparison

Table 2 shows the distribution of positive, negative and neutral responses among people from groups 1 and 2. People in group 2 express more satisfaction than in group 1. This corroborates the fact that we succeeded in transcribing future users' needs. This is the same for usefulness, benefits and usage intention, for which we collected more positive appreciations from the participants who were not involved in the design. This may be related to the surprise effect and let's expect a motivating effect for further use. The negative appreciations about usefulness in group 1 were given by participant P4 concerning quality indicators. This can be explained by the position of the participant (engineer) and his seniority. He stated that *"engineers use ALEX only in specific maintenance operation periods"*. As he is an expert, quality indicators do not have particular usefulness for him.

### 4.2.3 Comments of Participants Related to Each Criteria

The participants provided us with very valuable comments. A more complete transcription of these interviews can be found in (Touré *et al*, 2017). Here, we give the most salient comments.

**Usefulness.** Six out of ten participants explicitly found comments functions useful: four from group 2 and two from group 1. Three out of ten participants explicitly found indicators useful. One relates that: *"... in the previous version of ALEX, we couldn't really rely on sheets during maintenance operations... as the information evolves rapidly, when someone notices a mistake or something else... it was discussed face to face with the person supposed to operate the sheet's modifications ... which was done... or not... for me, comments are a feature more rewarding than oral exchanges, comments come from everyone... a trace of their viewpoint is kept, less chance of forgetting or losing information as was common in previous versions"*. Indeed, having up-to-date information is an important part of information quality, positively related to an effective use of the platform (DeLone and McLean, 2003) by two participants from group 2 and one from group 1. Concerning 'Likes', three out of ten participants explicitly found this functionality useful: one from group 2 and two from group 1. One participant qualified as *"sympathetic"* the idea of adding this social functionality and highlighted *"a lack of communication and social components in the previous version of ALEX"*.

**User satisfaction and benefits.** Participant generally find the platform more modern and satisfactory overall; they did not find many negative aspects. Participant P1 said that ALEX was *"a renovated tool, similar to those findable in the internet market, more playful and pleasant"*, while participant P4 argued that Alex was *"more user-friendly"*. They express benefits to use new 'comment' and 'Like' functionalities. Participant P5 employed the phrases *"peers' acknowledgment and feeling useful"*. Participant P3 said: *"...as it is now easier to use, we have more time to submit and seek for information... I am personally satisfied to participate in the building of the tool... inter alia to help the new colleagues integrating in the company... but I would like to be aware of my exact role in the tool and also have a kind of acknowledgements from the company..."* In these words, we identify the belief of social influence which arises from the use of the tool and motivates users. This is an interesting finding, as intrinsic benefits like reputation, joy and knowledge growth positively leverage continued use of knowledge-sharing tools (He & Wei, 2009).

**Impact on use and on performance.** At the time of the interviews, most participants did not mention any significant increase of ALEX use compared with how they used the previous versions. Four out of ten participants were frequent users (from twice a week to every day, according to the working tasks to perform), while the remainder used it once a month or less. When asked why, most of them answered that they had enough experience and knowledge of the hydraulics infrastructures. They also justified this by the fact that they were rarely confronted with difficult or atypical issues they didn't already know how to deal with, or needed more frequent connection to ALEX. Nevertheless, three participants argued that ALEX had been *"a time saver to access unknown intervention venues"* and useful to *"get information about the components of my new OC"*, or to assist him *"during a drain, a common maintenance operation"* in water infrastructures. The two first ones were new to the OC, and the last one had a complex maintenance operation to perform.

**Usage intention.** About half of the participants expressed usage intention linked with information seeking. Few of them plan to submit and collaborate on ALEX's content. However, the score 42.9% of neutral appreciation rate (see Table 1) can be explained be the youth of the project and the particular conditions of the context. For example, participant P2 was about to leave SCP (termination of his contract). He nevertheless participated in the evaluation and specified that *"ALEX usage perspectives are positive ... under the conditions of a general advertisement campaign within the company..."*. Participant P1 also stressed the positive effect of the user-centred design approach on workers' involvement and on sustainability of the new ALEX: *"... everyone participated in the refinement of the tool, the result satisfied more people and strengthened the project... everyone sees more clearly its real interest, which was not necessarily the case before, so I think it will be continuously used ..."*.

### 4.2.4 Conclusion about Evaluation

The qualitative evaluation we conducted showed that workers were satisfied with their new tool. We also noticed that new beliefs arose from the use of the tool, such as social reputation, usefulness and joy. Participants showed positive usage intention, especially for information seeking, which is a way of knowledge verification and learning. However, the usage frequency did not change, as workers considered themselves too experienced to change their habits. This is not surprising, as learners most of the time are poor at estimating their skills, but this can change by learning and improving their metacognitive skills (Glenberg, Wilkinson, & Epstein, 1982; Kruger & Dunning, 1999). We believe that this will have a positive impact on the tool usage in the long term. Further evaluation certainly needs to be done, but these outcomes corroborate the fact that our methodology plays a role in sustainable use, defended by our generic cycle of improvement of technologies. People engagement is supported half by involvement in the design methodology and half by the social functionalities that give positive beliefs and usage intention. Designers help users to express their latent needs and transcribe them; experts have roles as content validators and community moderators; and the other workers participate in the community. Results give us positive insights into the sustainability of our proposed model for informal learning.

## 5 DISCUSSIONS AND CONCLUSION

Our analysis showed that ESM are appropriate to support informal learning strategies in the workplace. Indeed, social features like comments, appreciations, activity indicators are adapted to stimulate use behaviors and support learning, particularly meta-cognitive aspects. Three adaptations must nevertheless be carried out: (1) Base the design on a precise and relatively exhaustive informational corpus of the procedures and know-how already formalized in the company and contextualize the access in the form of community of practice structured according to collaborative spaces; (2) Add indicators of judgment on the operational quality of information and the informational capital built, and (3) Define forms of moderation and control consistent with the hierarchical structures of the company. Our analysis also showed that an incremental and iterative approach of user-centered design had to be implemented to define how to adapt the design and to accompany change.

The reinforcement of the design work on information architectures, in terms of content, structuring and publication, is not contrary to the principle of social media. Evaluation shows that information seeking is a massive use intention. It thus could be useful to refine this work for proposing information search recommendation based on users' tracks. On the other hand, the need to adapt forms of moderation and control to the hierarchical structures of the company questions us. This principle is coherent with learning objectives since it creates some forms of mediation but is less so if one considers the principles of social media which consist in smoothing these forms of hierarchies to highlight the speech of each moderated by the collective. We wonder whether it is realistic to add this additional work load. Its implication is indeed critical to guarantee this type of functioning. In addition, we are wondering whether these requirements are indeed sustainable over the long term or whether they are an acceptance step in the design cycle as a form of temporary guaranty that should fall after the use of this type of platform all over the company.

The evaluation conducted shows promising results about uses and effects on the satisfaction and the feeling of learning after three months of use. On this basis, our next objectives will be to extend the deployment of the platform to all OCs to observe the acceptability of the principles to the whole organization, the informal learning effects and answer more general questions about the Forms of moderation.